\documentclass[11pt,a4paper]{article}
\usepackage{amsfonts}
\usepackage{amsmath}
\usepackage{amssymb}
\usepackage{amsthm}
\usepackage{graphicx}
\usepackage{booktabs}
\usepackage{theorem}
\usepackage{array}
\usepackage[numbers, sort]{natbib}
\usepackage[footnotesize]{caption}

\allowdisplaybreaks[1]

\numberwithin{equation}{section}

\newcommand{\be}{\begin{equation}}
\newcommand{\ee}{\end{equation}}
\newcommand{\beq}{\begin{equation}}
\newcommand{\eeq}{\end{equation}}
\newcommand{\bea}{\begin{eqnarray}}
\newcommand{\eea}{\end{eqnarray}}

\makeatletter

\newcommand{\Rmnum}[1]{\expandafter\@slowromancap\romannumeral #1@}
\makeatother

\begin{document}

\begin{titlepage}

\vspace*{-15mm}
\begin{flushright}
MPP-2010-52\\
\end{flushright}
\vspace*{0.7cm}

\begin{center}
{
\bf\LARGE
Measurable Neutrino Mass Scale in $\boldsymbol{A_4 \times SU(5)}$
}
\\[8mm]
S.~Antusch$^{\star}$
\footnote{E-mail: \texttt{antusch@mppmu.mpg.de}},
Stephen~F.~King$^{\dagger}$
\footnote{E-mail: \texttt{king@soton.ac.uk}},
M.~Spinrath$^{\star}$
\footnote{E-mail: \texttt{spinrath@mppmu.mpg.de}},
\\[1mm]

\end{center}
\vspace*{0.50cm}
\centerline{$^{\star}$ \it
Max-Planck-Institut f\"ur Physik (Werner-Heisenberg-Institut),}
\centerline{\it
F\"ohringer Ring 6, D-80805 M\"unchen, Germany}
\vspace*{0cm}
\centerline{$^{\dagger}$ \it
School of Physics and Astronomy, University of Southampton,}
\centerline{\it
SO17 1BJ Southampton, United Kingdom }
\vspace*{1.20cm}
\begin{abstract}

\noindent
We propose a supersymmetric $A_4\times SU(5)$ model of quasi-degenerate neutrinos
which predicts the effective neutrino mass $m_{ee}$ relevant for neutrinoless
double beta decay to be proportional to the neutrino mass scale,
thereby allowing its determination approximately independently of unknown 
Majorana phases.
Such a natural quasi-degeneracy is achieved by using $A_4$ family symmetry 
(as an example of a non-Abelian family symmetry with real triplet
representations) to enforce a contribution
to the neutrino mass matrix proportional to the identity.
Tri-bimaximal neutrino mixing as well as quark CP violation with $\alpha \approx 90^\circ$
and a leptonic CP phase $\delta_{\mathrm{MNS}} \approx 90^\circ$ arise from
the breaking of the $A_4$ family symmetry by the vacuum expectation values of four
``flavon'' fields pointing in specific postulated directions in flavour space.
\end{abstract}

\end{titlepage}

\setcounter{footnote}{0}

\section{Introduction}
Over the past dozen years or so our knowledge of the neutrino sector has increased dramatically with the discovery
of neutrino mass and mixing in atmospheric and solar neutrino oscillations, followed by the observation of
terrestrial neutrino oscillations in long baseline neutrino experiments which have confirmed and refined
the earlier results \cite{King:2007nw}. Yet, despite this progress, which may even be termed a ``neutrino revolution'',
there are many questions about neutrinos which remain unanswered. Perhaps the most pressing of these
is the origin, nature and magnitude of neutrino mass, since neutrino oscillations only provide
information about the squared mass differences between neutrino species which are independent of the
absolute neutrino mass scale or the nature of the neutrino mass (i.e. Dirac or Majorana).
In the absence of any confirmed
ex\-peri\-men\-tal signal from either beta decay end-point experiments or neutrinoless double beta decay
experiments, the most stringent limits on the absolute neutrino mass scale comes indirectly from
cosmology where one typically obtains the limit on the absolute neutrino mass scale expressed in terms
of the lightest neutrino mass as $m_{\mathrm{lightest}} \lesssim 0.2$ eV
\cite{Jarosik:2010iu}. Thus, there remains the interesting possibility that neutrinos are quasi-degenerate,
which one may roughly define as $m_{\mathrm{lightest}}>0.05$ eV, where the lower limit is approximately set equal to the square root of the
atmospheric neutrino mass squared difference.

The current generation of running or planned neutrinoless double beta decay experiments are capable of discovering
quasi-degenerate neutrinos, as defined above, within the next years. Such a discovery would herald a new neutrino
revolution to rival the last one, and would lead to an explosion of interest in theoretical models capable
of accounting for quasi-degenerate neutrinos. In general having quasi-degenerate neutrinos does
not lead to a sharp prediction for the neutrinoless double beta decay observable $m_{ee}$ as a function
of $m_1 \simeq m_2 \simeq m_3$ due to the presence of unknown phases in the neutrino mass matrix \cite{Feruglio:2002af}.
The general conclusion that unknown phases enter the prediction for $m_{ee}$
also remains valid in models which combine the experimental observation
of (at least approximate) tri-bimaximal (TB) \cite{Harrison:2003aw}
lepton mixing with the possibility of a quasi-degenerate neutrino mass
spectrum. The reason is that the relevant phases entering $m_{ee}$
are Majorana phases.
Allowing for arbitrary Majorana phases and considering a quasi-degenerate neutrino mass spectrum and TB mixing,
$m_{ee}$ can still be in the approximate interval 
$m_{ee} \in[m_{\mathrm{lightest}}/3,m_{\mathrm{lightest}}]$.

What would we learn about the origin of neutrino mass
from the discovery of quasi-degenerate neutrinos in neutrinoless double beta decay?
Clearly this would imply that neutrinos are Majorana, and possibly (but not necessarily) that would indicate that
a seesaw mechanism is at work, but what kind of seesaw mechanism, i.e. is it type I or II?\footnote{We shall not consider the type III or further types of seesaw mechanism in this paper.}
There are known examples of type I and type II seesaw models which can lead to quasi-degenerate neutrinos as
well as TB lepton mixing, so clearly quasi-degenerate neutrinos would not distinguish different types of seesaw
mechanism. For example, the supersymmetric (SUSY) Grand Unified Theory (GUT) based on $SO(10)$
with family symmetry $PSL(2,7)$ proposed in \cite{King:2009tj} is based on the type II seesaw mechanism and
leads to TB mixing and allows quasi-degenerate neutrinos. On the other hand the SUSY $A_4$ model in \cite{Altarelli:2005yx}
based on the type I seesaw mechanism also leads to TB mixing and allows quasi-degenerate neutrinos.
Interestingly, the SUSY
$A_4 \times SU(5)$ model with type I seesaw mechanism
does not favour quasi-degenerate neutrinos \cite{Altarelli:2008bg}, whereas a related model with
a type II seesaw mechanism does allow quasi-degenerate neutrinos
\cite{Ciafaloni:2009ub}.
More generally, there is a huge literature on family symmetry models based on $A_4$ \cite{Ma:2007wu}
or other symmetries \cite{Barbieri:1999km}. 
However, to our knowledge, in all above examples, quasi-degenerate neutrinos are subject to the mentioned
phase uncertainties in the prediction of $m_{ee}$ as a function of $m_1$.

It is interesting to ask, in what class of theories would we learn the most about the
neutrino mass scale $m_{\mathrm{lightest}}$ from the discovery of a measurement of $m_{ee}$ in
neutrinoless double beta decay? Clearly the answer would be those theories which predict
$m_{ee}$ uniquely as a function of $m_{\mathrm{lightest}}$ without ambiguities from unknown phases, but
the next question is do such theories exist? Perhaps surprisingly the answer is in the affirmative,
and, even more surprisingly, the class of theories which have this property turn out to suggest 
the way that the neutrino mass matrix is generated, namely by a usual type I seesaw contribution with two or three right-handed neutrinos, plus an additional contribution proportional to the unit matrix. 
In \cite{Antusch:2004xd}, two of us proposed
a class of theories of exactly this kind
which we referred to as a ``type II upgrade of type I seesaw models''.
In this class of models the additional contribution to the neutrino mass 
matrix was realised by an additional type II seesaw. 
The type I seesaw part of the neutrino mass matrix, which controls the
mass squared differences and mixing angles,
was governed by sequential right-handed neutrino dominance \cite{King:1998jw}.
The effect of such an additional unit matrix structure implies that for
quasi-degenerate neutrino masses the Majorana CP phases are small and thus
$m_{ee}\approx m_{\mathrm{lightest}}$.
Although the class of models was specified, no realistic ``type II upgrade''
model has ever been proposed.

In this paper we shall propose a model following the idea of an additional contribution the neutrino mass matrix proportional to the unit matrix based on $A_4$ family symmetry with $SU(5)$ grand unification.
The model contains tri-bimaximal neutrino mixing after the $A_4$ family symmetry is broken as an indirect result
of the assumed aligned ``neutrino flavons'' in the type I seesaw sector via constrained sequential dominance \cite{King:2005bj}.
These ``neutrino flavons'' break the $A_4$ symmetry, being assumed to be
aligned along the columns of the TB mixing matrix,
but quadratic combinations of the neutrino flavons respect accidentally the neutrino flavour symmetry as discussed
in \cite{King:2009ap}. Further ``quark flavons'' are assumed to be
misaligned compared to the ``neutrino flavons''
and are, together with the ``neutrino flavons'', responsible for quark and charged lepton masses and quark mixings.
As expected, due to a possibly large type II seesaw contribution, or alternatively due to an additional type I seesaw contribution from an additional triplet representation of right-handed neutrinos, the model can naturally predict the neutrinoless double beta decay mass observable to be
approximately equal to the neutrino mass scale.
We also make a detailed fit to quark masses and mixing using the misaligned quark flavons and show that a simple ansatz
for the phase of one of the misaligned quark flavons leads to successful quark CP violation.
In order for radiative corrections not to modify too much the TB mixing for quasi-degenerate
neutrinos \cite{Varzielas:2008jm},
we shall restrict ourselves to low values of the ratio of Higgs vacuum expectation values (vevs) $\tan \beta <1.5$.
For such low $\tan \beta < 1.5$, a viable GUT scale ratio of $y_\mu/y_s$ is achieved within SUSY $SU(5)$ GUTs using a Clebsch factor of $9/2$, as proposed recently by two of us in \cite{Antusch:2009gu}.
For the third generation we use $b$-$\tau$ Yukawa coupling unification $y_\tau/y_b = 1$ at the GUT scale which is viable for low $\tan \beta$ (see, e.g., \cite{Ross:2007az}).

The layout of the remainder of the paper is as follows.
In section 2 we present the model. In section 3 we perform a numerical fit to the
quark and charged lepton masses and quark mixing angles and CP violating phase and discuss the neutrino masses
and lepton mixing angles. Section 4 summarises and concludes the paper. In Appendix A we give a renormalisable superpotential and explicit expressions for the effective couplings and Appendix B contains a possible vacuum alignment.

\section{The Model}

In this section, we propose and describe a SUSY GUT model based on the unified $SU(5)$
gauge group as well as on the family symmetry $A_4$ amended by some discrete $\mathbb{Z}_2^2\times \mathbb{Z}_4^2$ symmetries and an $U(1)_R$ symmetry
 as specified in Tab.~\ref{Tab:Symmetries}.

\subsection{Symmetries and Field Content of the $\boldsymbol{SU(5)}$ GUT Model}

\begin{table}
\centering
\begin{tabular}{cccccccc}
\toprule
& $SU(5)$ & $A_4$ & $\mathbb{Z}_2$ & $\mathbb{Z}'_4$ & $\mathbb{Z}'_2$ & $\mathbb{Z}_4$ & $U(1)_R$\\
\midrule
\multicolumn{8}{l}{Chiral Matter}  \\
\midrule
$F$ &  $\mathbf{\overline{5}}$ & $\mathbf{3}$ & + & 0 & + & 0 & 1 \\
$T_1$, $T_2$, $T_3$  &  $\mathbf{10}$, $\mathbf{10}$, $\mathbf{10}$ &  $\mathbf{1}$, $\mathbf{1}$, $\mathbf{1}$ & +, +, - &  0, 1, 0 &  +, +, + &  1, 0, 0 & 1, 1, 1  \\
$N_1$, $N_2$ &  $\mathbf{1}$, $\mathbf{1}$ &  $\mathbf{1}$, $\mathbf{1}$ & +, + &  0, 1 &  +, + &  1, 0  & 1, 1  \\
\midrule
\multicolumn{8}{l}{Flavons \& Higgs Multiplets}  \\
\midrule
$\phi_{23}$, $\phi_{123}$, $\phi_{3}$ & $\mathbf{1}$ & $\mathbf{3}$  & +, +, - & 0, 3, 0 & +, + ,+ & 3, 0, 0 & 0, 0 ,0 \\
$\tilde{\phi}_{23}$ & $\mathbf{24}$ & $\mathbf{3}$  & + & 3 & - & 0   & 0\\
$H_5$, $\bar{H}_5$ &  $\mathbf{5}$, $\mathbf{\overline{5}}$ & $\mathbf{1}$, $\mathbf{1}$ & +, + & 0, 0  & +, + & 0, 0 & 0, 0\\
$H_{15}$, $\bar{H}_{15}$ &  $\mathbf{15}$, $\mathbf{\overline{15}}$ &  $\mathbf{1}$, $\mathbf{1}$  & +, + &  0, 0 &  +, + & 0, 0 & 0, 0\\
$H_{45}$, $\bar{H}_{45}$ &  $\mathbf{45}$, $\mathbf{\overline{45}}$ &  $\mathbf{1}$, $\mathbf{1}$  & +, + &  0, 0 &  -, - & 0, 0 & 0, 0\\
\midrule
\multicolumn{8}{l}{Matter-like Messengers } \\
\midrule
$A_5$, $\bar{A}_5$ &  $\mathbf{5}$, $\mathbf{\overline{5}}$ & $\mathbf{1}$, $\mathbf{1}$ & +, + &   1, 3 & -, - &  0, 0 & 1, 1 \\
$A_{10}$, $\bar{A}_{10}$ &  $\mathbf{10}$, $\mathbf{\overline{10}}$ & $\mathbf{3}$, $\mathbf{3}$ & +, + & 0, 0 &  +, + &  0, 0   & 1, 1 \\
$A_{1}$ &  $\mathbf{1}$ & $\mathbf{3}$ & + &   0 &  + &  0   & 1\\
\midrule
\multicolumn{8}{l}{Higgs-like Messengers }  \\
\midrule
$B$, $\bar{B}$ & $\mathbf{5}$, $\mathbf{\overline{5}}$  & $\mathbf{1}$, $\mathbf{1}$ & +, + & 2, 2 & +, + &  0, 0  & 0, 2  \\
$C_1$, $\bar{C}_1$ & $\mathbf{1}$, $\mathbf{1}$  & $\mathbf{1}$, $\mathbf{1}$ & +, + &  0, 0  & +, + &  2, 2 & 2, 0  \\
$C_2$, $\bar{C}_2$ & $\mathbf{1}$, $\mathbf{1}$  & $\mathbf{1}$, $\mathbf{1}$ & +, + &  2, 2  & +, + &  0, 0 & 2, 0  \\
\bottomrule
\end{tabular}
\caption{Representations and charges of the superfields. The subscript $i$ on the fields $T_i$, $N_i$ and $C_i$ is a family index. The flavon fields $\phi_i$, $\tilde{\phi}_{23}$ can be associated to a family via their charges under $\mathbb{Z}_2^2\times \mathbb{Z}_4^2$.  The subscripts on the Higgs fields $H$, $\bar{H}$ and extra vector-like matter fields $A$, $\bar{A}$ denote the transformation properties under $SU(5)$. \label{Tab:Symmetries}}
\end{table}

Let us start introducing the model by specifying the field content and the symmetries.
The Standard Model matter fields fit nicely into the two representations $\overline{\mathbf{5}}$, which we call $F$, and $\mathbf{10}$, which we call $T$. Explicitly they are given as
\begin{equation}
\begin{split}
F_i =  &\begin{pmatrix}
       d_R^{c} & d_B^{c} & d_G^{c} & e &-\nu
        \end{pmatrix}_i \:, \\
T_i = \frac{1}{\sqrt{2}}
         &\begin{pmatrix}
         0 & -u_G^{c} & u_B^{c} & -u_{R} & -d_{R} \\
         u_G^{c} & 0 & -u_R^{c} & -u_{B} & -d_{B} \\
         -u_B^{c} & u_R^{c} & 0 & -u_{G} & -d_{G} \\
         u_{R} & u_{B} & u_{G} & 0 & -e^c \\
         d_{R} & d_{B} & d_{G} & e^c & 0
         \end{pmatrix}_i \:,
\end{split}
\end{equation}
where the lower indices $R$, $B$ and $G$ denote the quark colours and $i = 1,2,3$ is the family index. In our model, we consider that the three generations $F_i$ form a triplet representation ${\bf 3}$ of an $A_4$ family symmetry whereas the three generations $T_i$ form singlets $\mathbf{1}$ under $A_4$.\footnote{We note that in principle any non-Abelian family symmetry with real triplet
representations, like, e.g., $SO(3)$, would in principle be suitable for the construction of models with additional contributions to the neutrino mass matrix proportional to the unit matrix. In this paper we focus on $A_4$ as a specific example.} In the following, we suppress the $A_4$ indices. In addition, we consider two right-handed neutrinos, singlets under SU(5) as well as under $A_4$, labeled by $N_1$ and $N_2$.

We furthermore consider fifteen-dimensional Higgs representations $H_{15}$, $\bar{H}_{15}$ which contain $SU(2)_L$-triplet Higgs fields that obtain induced vevs after electroweak symmetry breaking. $H_{15}$ induces in this way a type II seesaw contribution to the neutrino mass matrix which is, to leading order, proportional to the unit matrix and can increase the neutrino mass scale without modifying the values for the leptonic mixing angles.

$SU(5)$ is spontaneously broken by the vev of the $\tilde{\phi}_{23}$ field, electroweak symmetry is broken by the vevs of the Higgs fields $H_5$, $\bar H_5$, $H_{45}$, $\bar H_{45}$ and $A_4$ is spontaneously broken by the vevs of the flavon fields, i.e.\ the family symmetry breaking Higgs fields $\phi_{123}$, $\phi_{23}$, $\phi_{3}$ and  $\tilde{\phi}_{23}$. We comment below on the specific directions in which we assume $A_4$ to be broken by the flavons.

On top of that, we consider additional ``messenger'' fields which are heavy and which, after effectively integrating them out of the theory, give rise to higher-dimensional operators generating the Yukawa coupling matrices as well as the mass matrix of the gauge singlet (right-handed) neutrinos $N_i$.

The field content of our model as well as the symmetries are specified in Tab.~\ref{Tab:Symmetries}. We note that it is always possible to replace any product of commuting discrete symmetries by a single Abelian group $U(1)$
with a suitable choice of charges for the fields, so it is possible to replace the
$\mathbb{Z}_2^2\times \mathbb{Z}_4^2$ symmetry by a single $U(1)$ symmetry, with an appropriate choice of charges.
Indeed many models in the literature use an Abelian $U(1)$ symmetry rather than a product
of $\mathbb{Z}_N$ symmetries to control the operators. Although this looks simpler, it should be remarked that firstly an Abelian
symmetry has infinitely many more group elements than any discrete symmetry, and secondly one must then confront the
question of Goldstone bosons once the assumed global Abelian symmetry is broken.
If the Abelian symmetry is gauged one must further
complicate the model by ensuring that it is anomaly free. Therefore an auxiliary discrete symmetry, even a large one, has definite advantages
over an Abelian symmetry. Furthermore discrete symmetries are ubiquitous in string theory constructions.
Finally, the auxiliary discrete symmetry used here is rather a simple one consisting of a product of $\mathbb{Z}_2$ and $\mathbb{Z}_4$ parity
factors. Thus we regard the use of the discrete $\mathbb{Z}_2$ and $\mathbb{Z}_4$ symmetries as being a well motivated, simple and attractive alternative to the use of an Abelian $U(1)$ symmetry.

We would like to remark that we do not explicitly consider the full flavour and GUT Higgs sector of the model and just assume that the $SU(5)$ and $A_4$ breaking vevs are aligned in the desired directions of field space. We assume that in these sectors issues like doublet-triplet splitting are resolved. Without specifying these sectors, a reliable calculation of the proton decay rate must also be beyond the scope of the present paper. The focus of the present paper is thus to illustrate that quasi-degenerate light neutrino masses can be realised together with a type II seesaw in a $SU(5)$ GUT framework.

We would furthermore like to remark that in addition to the type II seesaw contribution there is a possible additional contribution to the neutrino mass matrix proportional to the unit matrix from the messenger field $A_1$ which is a singlet under $SU(5)$ and a triplet under
$A_4$. When it is integrated out, it also induces a contribution to the neutrino mass operator which is proportional to the unit matrix.

\subsection{The Effective $\boldsymbol{A_4\times SU(5)}$ Symmetric Superpotential}

\begin{figure}
\centering
\includegraphics[scale=0.8]{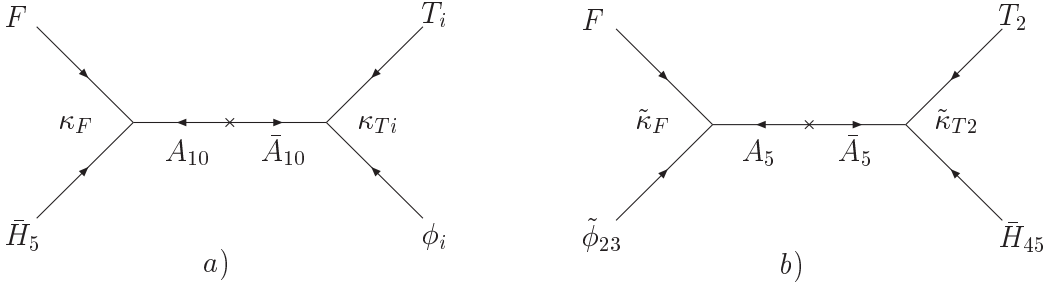}
\caption{Supergraph diagrams inducing the effective superpotential operators for the down-type quarks and charged leptons. \label{Fig:messenger_dl}}
\end{figure}

\begin{figure}
\centering
\includegraphics[scale=0.7]{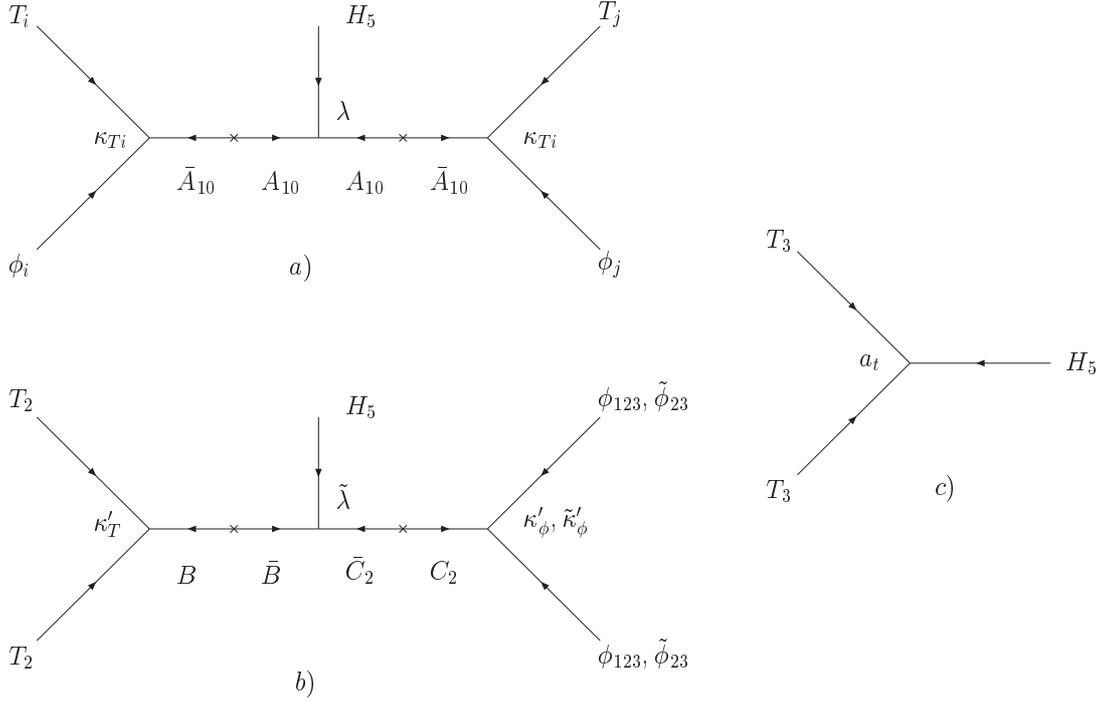}
\caption{Supergraph diagrams inducing the effective superpotential operators for the up-type quarks. \label{Fig:messenger_u}}
\end{figure}

\begin{figure}
\centering
\includegraphics[scale=0.70]{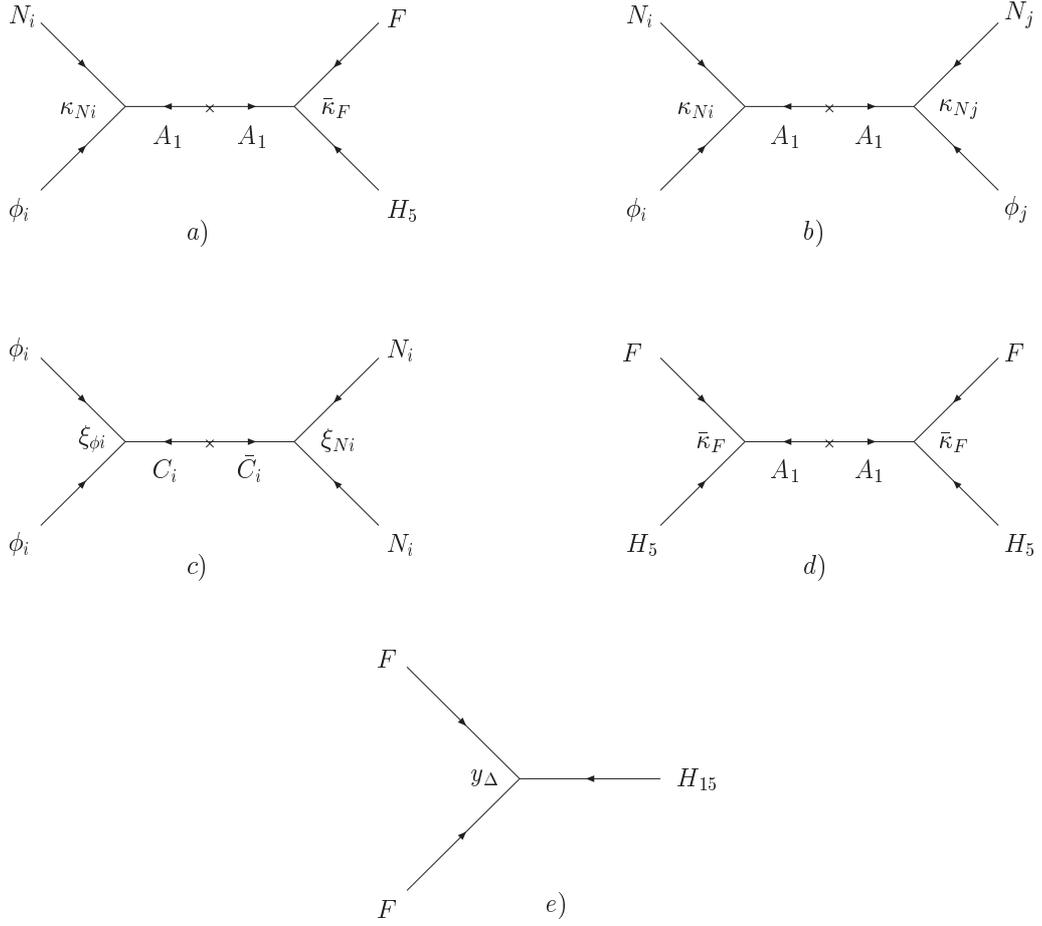}
\caption{Supergraph diagrams inducing the effective superpotential operators for the neutrino sector. \label{Fig:messenger_n}}
\end{figure}

The renormalisable superpotential resulting from Tab.~\ref{Tab:Symmetries} is given in the Appendix.
Integrating out the heavy messenger superfields denoted by $A$, $B$ and $C$,
the Feynman diagrams in Figs.~\ref{Fig:messenger_dl}, \ref{Fig:messenger_u} and \ref{Fig:messenger_n}
then lead to the effective non-renormalisable superpotential terms in the $SU(5)$ and $A_4$ unbroken phase:
\begin{align}
W_{Y_l} &= \frac{\sqrt{2}}{M_{A_{10}}} F \left( a_1 \phi_{23} T_1 + a_2 \phi_{123} T_2 + a_3 \phi_{3} T_3 \right) \bar{H}_5 + \frac{\sqrt{2} \, \tilde{a}_2}{M_{A_5}} F \tilde{\phi}_{23} T_2 \bar{H}_{45} \:,\label{Eq:Yl} \\
W_{Y_u} &= \frac{1}{4} \left( \frac{a_{12}}{M_{A_{10}}^2} T_1 T_2 (\phi_{123} \cdot \phi_{23}) + \frac{a_{13}}{M_{A_{10}}^2} T_1 T_3 (\phi_3 \cdot \phi_{23}) + \frac{a_{23}}{M_{A_{10}}^2} T_2 T_3 (\phi_{123} \cdot \phi_3)    \right) H_5\:, \nonumber\\
& + \frac{1}{4} \left( a_{33} T_3^2 + \frac{a_{22}}{M_{A_{10}}^2} T_2^2 \phi_{123}^2 + \frac{a_{11}}{M_{A_{10}}^2} T_1^2 \phi_{23}^2 + \frac{\tilde{a}_{22}}{M_{A_5}^2} T_2^2 \tilde{\phi}_{23}^2   \right) H_5 \label{Eq:Yu} \:,\\
W_{Y_\nu} &= \frac{1}{M_{A_{10}}} F \left( a_{\nu_1} \phi_{23} N_1 + a_{\nu_2} \phi_{123} N_2 \right) H_5 \:, \label{Eq:Ynu} \\
W_{\nu}^{\Delta} &= y_\Delta H_{15} F F\:, \label{Eq:Delta} \\
W_{\nu}^{d=5} &= \frac{\bar{\kappa}_F^2}{M_{A_1}} F H_5 F H_5 \:, \label{Eq:d=5} \\
W_{\nu}^{M_R} &= \frac{a_{R_{11}}}{M_{A_{10}}^2} \phi_{23}^2 N_1^2 + \frac{a_{R_{22}}}{M_{A_{10}}^2} \phi_{123}^2 N_2^2 + \frac{a_{R_{12}}}{M_{A_{10}}^2} \left( \phi_{123} \cdot \phi_{23} \right) N_1 N_2 \:.\label{Eq:MR}
\end{align}
After GUT symmetry breaking the $SU(2)_L$ doublet components from $H_5$ and $H_{45}$ respectively $\bar{H}_5$ and $\bar{H}_{45}$ mix and only the light states acquire the $SU(2)_L$ breaking vevs which give the fermion masses, as discussed in Appendix A where the effective couplings $a$ appearing in the effective superpotential are also explicitly given.

\subsection{Assumed Vacuum Alignment}

In the following we assume that the vevs of the $A_4$ breaking flavon fields point in the following directions in field space such that
\begin{equation} \label{Eq:vacuumalignment}
b_1 \frac{\langle \phi_{23} \rangle }{M_{A_{10}}}  = \begin{pmatrix}  0 \\  1\\  - 1  \end{pmatrix} \epsilon_{23} \;, \quad
b_2 \frac{\langle \phi_{123} \rangle }{M_{A_{10}}}  = \begin{pmatrix} 1 \\  1 \\ 1  \end{pmatrix} \epsilon_{123} \;, \quad
b_3 \frac{\langle \phi_{3} \rangle }{M_{A_{10}}}  = \begin{pmatrix}  0 \\  0\\  1  \end{pmatrix} \epsilon_{3} \;.
\end{equation}
The $b_i$ are defined in Appendix A. These relations also define the quantities $\epsilon_{123}$, $\epsilon_{23}$ and $\epsilon_{3}$. The breaking of $A_4$ along the field directions of  $\phi_{123}$ and $\phi_{23}$ allows us to realise tri-bimaximal neutrino mixing via constrained sequential dominance (CSD) \cite{King:1998jw}. It is also worth noting that the flavon vevs $\langle \phi_{123} \rangle$ and $\langle \phi_{23} \rangle$
are orthogonal, causing some of the terms in the superpotential to give a vanishing contribution to the mass matrices.
In the following, we assume that CP is only broken spontaneously by the vev of the flavon $\tilde{\phi}_{23}$.

For the flavon $\tilde{\phi}_{23}$ one may suppose {\it a priori} a less constrained alignment,
\begin{equation} \label{Eq:vacuumalignment2}
\tilde{b}_2 \frac{\langle \tilde\phi_{23} \rangle }{M_{A_5}}  = \begin{pmatrix}  0 \\  v \\  w  \end{pmatrix} \tilde{\epsilon}_{23} \;.
\end{equation}
However {\it  empirically} we find that the numerical fit to quark masses and mixings, in
particular quark CP violation, seems strongly to prefer that
the vacuum alignment of the flavon $\tilde{\phi}_{23}$ has its second component along the imaginary direction.
To simplify the results of the numerical fit we shall restrict ourselves to the case:
\begin{equation}
v = - \mathrm{i} \;.
\label{ansatz}
\end{equation}
In some future more ambitious theory one may attempt to reproduce Eq.~\eqref{ansatz} as a result of some
special vacuum alignment, but here we shall simply regard it as a special choice,
or ansatz, which leads to a successful fit to quark CP violation.

In Appendix B, we will discuss another possibility, namely to realise the flavon  $\tilde{\phi}_{23}$ effectively by splitting it up into two flavons $\tilde{\phi}_{2}$ and $\tilde{\phi}_{3}$, where one gets a purely real and the other a purely imaginary vev. 
For the effective superpotential in Eqs.~(\ref{Eq:Yl}) - (\ref{Eq:MR}) this would correspond to simply replacing $\tilde{\phi}_{23}^2 \to \tilde{\phi}_{2}^2 + \tilde{\phi}_{3}^2$ and $\tilde{\phi}_{23} \bar H_{45} \to \tilde{\phi}_{2} \bar H_{45}  + \tilde{\phi}_{3} \bar H'_{5}$, with an additional Higgs field $H'_{5}$. The field content of this extended version of the model that now includes also a vacuum alignment sector, is presented in Appendix B in Tab.\ \ref{Tab:Symmetries2}. The predictions of the two model variants are identical at the level of precision discussed here.

\section{Numerical Fit to Fermion Masses and Mixings}

\subsection{The Quark and Charged Lepton Sector}

We define our conventions for the Yukawa matrices such that the operators of the form $F T \phi \bar{H}$ and $T^2 \phi^2 H$ give the following Yukawa terms in the Lagrangian:
\begin{equation}
\mathcal{L}_{\mathrm{Yuk}} = - (Y^*_d)_{ij} Q_i \bar{d}_j H_d - (Y^*_e)_{ij} L_i \bar{e}_j H_d  - (Y_u^*)_{ij} Q_i \bar{u}_j H_u + \mathrm{h.c.}\;,
\end{equation}
where the $SU(5)$ relation $Y_d = Y_e^T$ would be fulfilled, if all Clebsch--Gordan factors were one. The convention we use here is the same as the one used by the Particle Data Group \cite{Amsler:2008zzb}.

From Eqs.~\eqref{Eq:Yl}, \eqref{Eq:Yu}, \eqref{Eq:vacuumalignment}, \eqref{Eq:vacuumalignment2} and \eqref{ansatz} the Yukawa matrix coupling the up-type quarks to the light up-type Higgs doublet with the $b$ coefficients as defined in Appendix A is given as
\begin{align}
Y_u &= \begin{pmatrix} 2 b_{11} \epsilon^2_{23} & 0 &  b_{13} \epsilon_{23} \epsilon_3 \\ 0 & 3 b_{22} \epsilon^2_{123} +  (w^2-1) \tilde{b}_{22}^2 \tilde{\epsilon}^2_{23} & b_{23} \epsilon_{123} \epsilon_3 \\  b_{13} \epsilon_{23} \epsilon_3 &  b_{23} \epsilon_{123} \epsilon_{3} & b_{33}  \end{pmatrix}, \label{Eq:GUTYu}
\end{align}
whereas the Yukawa matrices coupling the down-type quarks and charged leptons to the light down-type Higgs doublet are given as
\begin{align}
Y_d &= \begin{pmatrix} 0 & \epsilon_{23} & - \epsilon_{23} \\ \epsilon_{123} & \epsilon_{123} + \mathrm{i} \, \tilde{\epsilon}_{23} & \epsilon_{123} + w \tilde{\epsilon}_{23} \\ 0 & 0 & \epsilon_{3} \end{pmatrix}, \label{Eq:GUTYd}\\
Y_e^T &= \begin{pmatrix} 0 & c_{23} \epsilon_{23} & - c_{23} \epsilon_{23} \\ c_{123} \epsilon_{123} & c_{123} \epsilon_{123} + \mathrm{i} \, \tilde{c}_{23} \tilde{\epsilon}_{23} & c_{123} \epsilon_{123} + w \tilde{c}_{23} \tilde{\epsilon}_{23} \\ 0 & 0 & c_3 \epsilon_{3} \end{pmatrix} , \label{Eq:GUTYe}
\end{align}
where $c_{3}$, $c_{23}$, $\tilde c_{23}$ and $c_{123}$ are the Clebsch--Gordan factors arising from GUT symmetry breaking, see, e.g., \cite{Antusch:2009gu}. We have used the orthogonality of $\langle \phi_{23} \rangle$ and $\langle \phi_{123} \rangle$ and considered the above described notation for the flavon vevs. We note that in the definition for the Yukawa matrices we have introduced a complex conjugation why here appears a phase factor of $+\mathrm{i}$ in the 2-2 elements of the down-type quark and charged lepton Yukawa matrices.

With the given representations of the flavons, we obtain the following Clebsch--Gordan coefficients
\begin{equation}
\quad c_{123} = 1 \;,  \quad c_{23} = 1 \;, \quad c_3 = 1 \;, \quad \tilde{c}_{23} = 9/2 \;.
\end{equation}
For small values of $\tan \beta$ as we consider, the 1-loop SUSY threshold corrections are small and, taking the actual  experimental values of the strange quark and muon masses into account, the GUT scale value of $y_\mu / y_s$ prefers $\tilde{c}_{23} = 9/2$, as argued in \cite{Antusch:2009gu} (see also \cite{Antusch:2008tf}).

From the charged lepton Yukawa matrix we can derive the following approximate relations for the eigenvalues
\begin{equation}
y_\tau = c_3 \epsilon_3  \;, \quad y_\mu = \vert c_{123} \epsilon_{123} + \mathrm{i} \, \tilde{c}_{23} \tilde{\epsilon}_{23} \vert \;, \quad y_e = \frac{c_{23} \epsilon_{23} c_{123} \epsilon_{123} }{y_\mu } \;.
\end{equation}
Furthermore, since there is no 1-2 mixing from the up-sector, the mixing angle $\theta_{12}$ is approximately given as
\begin{equation}
\theta^{\mathrm{CKM}}_{12} = \left\vert \frac{\epsilon_{23} }{\epsilon_{123} + \mathrm{i} \, \tilde{\epsilon}_{23}} \right\vert \;.
\end{equation}
From those four equations the four $\epsilon$'s can be calculated and the relation for the CKM phase gives at the GUT scale
\begin{equation}
\vert \tan \delta_{\mathrm{CKM}} \vert = \left\vert \frac{\tilde{\epsilon}_{23}}{\epsilon_{123}} \right\vert \approx 1.22 \;. \label{Eq:deltaCKM}
\end{equation}
The RG evolution of the measured value for $\delta_{\mathrm{CKM}}$ gives a GUT scale value of $1.20$. So the value for the CKM phase (based on our assumed vacuum alignment) is already remarkably good if we only take the lepton masses and the value for $\theta_{12}$ into account which are measured to  high accuracy.\footnote{We would like to remark that with the assumed spontaneous CP violation, real $\det Y_u$ and $\det Y_d$ and with the small $|\tilde{\epsilon}_{23}| = {\cal O}(10^{-4})$, the model might also provide a solution to the strong CP problem, along the lines discussed in \cite{Nelson:1983zb}.}

\begin{table}
\begin{center}
\begin{tabular}{cc}
\toprule
Parameter & Value \\ \midrule
$2 b_{11} \epsilon_{23}^2 $ in $10^{-6}$ & $9.62$ \\
$3 b_{22} \epsilon_{123}^2$ in $10^{-4}$ & $-1.10$ \\
$(w^2 - 1) \tilde{b}_{22} \tilde{\epsilon}_{23}^2$ in $10^{-3}$ & $-1.10$ \\
$b_{13} \epsilon_{23} \epsilon_3$ in $10^{-3}$  & $-2.92$ \\
$b_{23} \epsilon_{123} \epsilon_3$ in $10^{-2}$ & $3.21$ \\
$b_{33}$ & $2.44$ \\ \midrule
$\epsilon_{123}$ in $10^{-5}$ & $5.88$ \\
$\epsilon_{23}$ in $10^{-5}$ & $4.30$ \\
$\tilde{\epsilon}_{23}$ in $10^{-4}$ & $-1.61$ \\
$\epsilon_{3}$ in $10^{-2}$ & $1.12$ \\
$w$ & $1.44$ \\
\bottomrule
\end{tabular}
\caption{The model parameters for $\tan \beta = 1.4$ and $M_{\mathrm{SUSY}} = 500$ GeV from a fit to the experimental data.
\label{Tab:Parameters}}
\end{center}
\end{table}

\begin{table}
\begin{center}
\begin{tabular}{cccc}
\toprule
Quantity (at $m_t(m_t)$) & Model & Experiment & Deviation  \\ \midrule
$y_\tau$ in $10^{-2}$ & 1.00 & 1.00 & $-0.027$\% \\
$y_\mu$ in $10^{-4}$  & 5.89 & 5.89 & $-0.029$\% \\
$y_e$ in $10^{-6}$ & 2.79 & 2.79 & $-0.130$\% \\  \midrule
$y_b$ in $10^{-2}$ & 1.58 & $1.58 \pm 0.05$ & $0.086 \sigma $  \\
$y_s$ in $10^{-4}$ & 2.83 & $2.99 \pm 0.86$ & $- 0.184 \sigma $ \\[0.1pc]
$y_d$ in $10^{-6}$ & 27.6 & $15.9^{+6.8}_{-6.6}$ & $1.723 \sigma$   \\  \midrule
$y_t$ & 0.938 & $0.936 \pm 0.016$ & $0.084 \sigma $   \\
$y_c$ in $10^{-3}$ & 3.54 & $3.39 \pm 0.46$ & $0.318 \sigma$   \\[0.1pc]
$y_u$ in $10^{-6}$ & 6.70 & $7.01^{+2.76}_{-2.30}$ & $- 0.134 \sigma$  \\ \midrule
$\theta_{12}^{\mathrm{CKM}}$ & 0.2257 & $0.2257^{+0.0009}_{-0.0010}$ & $- 0.022\sigma $  \\[0.3pc]
$\theta_{23}^{\mathrm{CKM}}$ & 0.0413 & $0.0415^{+0.0011}_{-0.0012}$ & $0.004 \sigma $ \\[0.1pc]
$\theta_{13}^{\mathrm{CKM}}$ & 0.0036 & $0.0036 \pm 0.0002$  & $-0.157\sigma $  \\[0.1pc]
$\delta_{\mathrm{CKM}}$ & 1.1782 & $1.2023^{+0.0786}_{-0.0431}$  & $-0.560 \sigma $  \\
\bottomrule
\end{tabular}
\caption{Fit results for the quark Yukawa couplings and mixing and the charged lepton Yukawa couplings at low energy compared to experimental data. A pictorial representation of the agreement between our fit and experiment can be also found in Fig.~\ref{Fig:FitResultsPlot}. \label{Tab:FitResults}}
\end{center}
\end{table}

For the detailed fit of the model to the data we applied the following procedure: We have taken the GUT scale Yukawa matrices from Eqs.~\eqref{Eq:GUTYu}, \eqref{Eq:GUTYd} and \eqref{Eq:GUTYe} and calculated their RG evolution down to the scale $m_t(m_t)$ for $\tan \beta = 1.4$\footnote{We note that in the MSSM small values of $\tan \beta$ are somewhat constrained due to bounds on the Higgs mass. However, we emphasize that our model may well be formulated in the context of the NMSSM or other non-minimal SUSY models where $\tan \beta$ of order one can readily be realised without these constraints.} and $M_{\mathrm{SUSY}} = 500$~GeV with the REAP software package \cite{Antusch:2005gp}. At the low scale we performed a $\chi^2$ fit to the quark masses and mixing and charged lepton masses depending on the parameters of the GUT scale Yukawa matrices. The fit gave a total $\chi^2$ of about 3.5 where we have assumed a relative error of 1~\% for the charged lepton masses and for the other observables we have taken the experimental errors.  Since we have 11 parameters and 13 observables this corresponds to a $\chi^2/\mathrm{dof}$ of about 1.6. This is a good fit since we have neglected theoretical uncertainties like, e.g., threshold corrections which could be treated as additional errors on the data lowering the total $\chi^2$.

\begin{figure}
\centering
\includegraphics[scale=0.7]{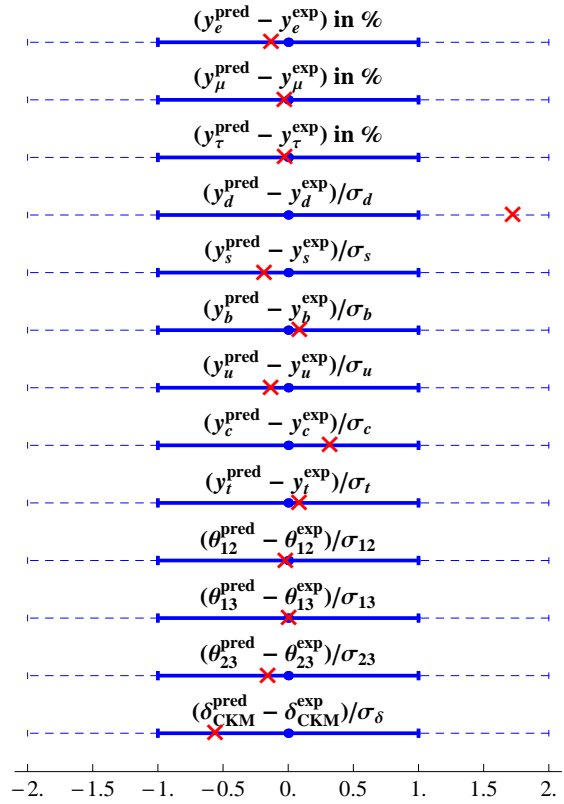}
\caption{Pictorial representation of the deviation of our fit from low energy experimental data for the charged lepton Yukawa couplings and quark Yukawa couplings and mixing parameters.  The deviations of the charged lepton masses are given in~\% while all other deviations are given in units of standard deviations $\sigma$. The straight blue lines give the 1~\% (1$\sigma$)  bound while the dashed lines give the 2~\% (2$\sigma$) bound. The red crosses denote our fit results. \label{Fig:FitResultsPlot} }
\end{figure}

The results for the GUT scale parameters are listed in Tab.~\ref{Tab:Parameters}. We would like to remark that these parameters depend on $\tan \beta$ and $M_{\mathrm{SUSY}}$ and also are subject to several theoretical uncertainties. For example, we note that the Higgs fields $H_{15}$ and $\bar{H}_{15}$ containing the Higgs triplets of the type II seesaw mechanism have masses at intermediate energy scale between $M_{\mathrm{GUT}}$ and $M_{\mathrm{EW}}$. Their effects are not included in the RG analysis. The effects are small and may be neglected if $y_{\Delta}$ is small, but they could be sizable if $y_{\Delta}$ is large.\footnote{Since the coupling $y_{\Delta}$ gives a contribution proportional to the unit matrix, it affects only  the RG evolution of the mass eigenvalues, but not of the mixing angles. Nevertheless, the possibility of additional RG effects from $y_{\Delta}$ provides a theoretical uncertainty in our setup.} Due to the additional theoretical uncertainties we do not explicitly give the errors on the high energy parameters or low energy fit results. The important input parameters for us are the charged lepton masses and quark mixing angles which have a  experimental error much smaller than these uncertainties.

In Tab.~\ref{Tab:FitResults} the low energy results are shown and compared to experimental data. A graphical illustration is given in Fig.~\ref{Fig:FitResultsPlot}. They illustrate that our minimal example model, with the assumed vacuum alignment of Eq.~\eqref{Eq:vacuumalignment} and \eqref{Eq:vacuumalignment2}, can fit well the data. We turn now to the results for the neutrino sector.

\subsection{The Neutrino Sector}

The neutrino Yukawa matrix is obtained from Eq.~\eqref{Eq:Ynu} as
\begin{align}
Y_{\nu} &= \begin{pmatrix} 0 &  b_{\nu_2} \epsilon_{123} \\  b_{\nu_1} \epsilon_{23} &  b_{\nu_2} \epsilon_{123} \\ -  b_{\nu_1} \epsilon_{23} &  b_{\nu_2} \epsilon_{123} \end{pmatrix} \;.
\end{align}
Additionally we have a diagonal mass matrix for the two right-handed neutrinos from Eq.~\eqref{Eq:MR},
\begin{equation}
M_R = \begin{pmatrix} 2 b_{R_1} \epsilon_{23}^2 & 0 \\ 0 & 3 b_{R_2} \epsilon_{123}^2  \end{pmatrix} \;,
\end{equation}
and contribution proportional to the unit matrix coming from Eqs.~\eqref{Eq:Delta} and \eqref{Eq:d=5},
\begin{equation}
M_L = \begin{pmatrix} m_0 & 0 &0 \\ 0 & m_0 & 0 \\ 0 & 0 & m_0 \end{pmatrix} \;.
\end{equation}

Using the seesaw relation
\begin{equation}
m_\nu = M_L - v_{u}^2 Y_\nu M_R^{-1} Y_\nu^T \;,
\end{equation}
we obtain for the neutrino mass matrix
\begin{equation}
m_\nu =
m_0 \begin{pmatrix} 1 & 0 &0 \\ 0 & 1 & 0 \\ 0 & 0 & 1 \end{pmatrix}
+
\frac{m^{I}_2}{3} \begin{pmatrix} 1 & 1 &1 \\ 1 & 1 & 1 \\ 1 & 1 & 1 \end{pmatrix}
+
\frac{m^{I}_3}{2} \begin{pmatrix} 0 & 0 &0 \\ 0 & 1 & -1 \\ 0 & -1 & 1 \end{pmatrix} \;
\end{equation}
with
\begin{equation}
m_0 = c_\gamma^2 v_u^2 \frac{y_{\Delta} \bar{\lambda}_{15}}{\mu_{15}} + c_\gamma^2 v_u^2 \frac{\bar{\kappa}_F^2}{M_{A_1}} \;, \quad
m^{I}_2 = -  v_u^2 \frac{b_{\nu_2}^2}{b_{R_2}}  \quad \text{and} \quad
m^{I}_3 = -  v_u^2 \frac{b_{\nu_1}^2}{b_{R_1}}  \;.
\end{equation}
In our model we therefore identify the neutrino masses as 
$m_1=m_0$, $m_2=m_0+m_2^I$, $m_3=m_0+m_3^I$, where without loss of generality
we can take $m_0$ to be positive and real while $m_2^I,m_3^I$ are real but can take 
either sign.
With $\vert m_0 \vert \gg \vert m_3^I \vert$,~$\vert m_2^I \vert$ a quasi-degenerate mass spectrum of the light neutrinos can be explained in a natural way. In the following, we will mainly restrict ourselves to this case.

From the structure of these matrices we see that we obtain tri-bimaximal mixing in the neutrino sector
\begin{eqnarray}
&&\theta_{13}^\nu = 0 \quad,\quad \theta_{23}^\nu = 45^\circ \quad, \quad \theta_{12}^\nu = \arcsin \frac{1}{\sqrt{3}} \approx 35.3^\circ  \:.
\end{eqnarray}
From the lepton sector we get the additional mixing contributions
\begin{eqnarray}
&&\theta_{13}^e = 0 \quad,\quad \theta_{23}^e = 0 \quad, \quad \vert \theta_{12}^e \vert  = \left\vert \frac{c_{123} \epsilon_{123}}{c_{123} \epsilon_{123} - \mathrm{i} \tilde{c}_{23} \tilde{\epsilon}_{23} } \right\vert \approx 4.6^\circ \;.
\end{eqnarray}
There is also a complex phase introduced by the charged lepton Yukawa matrix which can be calculated in the same way as in the quark sector
\begin{eqnarray}
\delta_{12}^e =  \arctan \frac{\tilde{c}_{23} \tilde{\epsilon}_{23} }{c_{123} \epsilon_{123}} \approx  - 85.4^\circ \label{Eq:deltae} \;.
\end{eqnarray}

For the approximate calculation of the MNS mixing parameters at the GUT scale we can use \cite{Antusch:2008yc}:
\begin{equation}
\begin{split}
s^{\mathrm{MNS}}_{23} &\approx s_{23}^\nu - \theta_{23}^e \\
s^{\mathrm{MNS}}_{13} \mathrm{e}^{- \mathrm{i} \delta^{\mathrm{MNS}}_{13} } &\approx \theta_{13}^\nu - s_{23}^\nu \theta_{12}^e \mathrm{e}^{- \mathrm{i} \delta_{12}^e}  \\
s^{\mathrm{MNS}}_{12} \mathrm{e}^{- \mathrm{i} \delta^{\mathrm{MNS}}_{12} } &\approx s_{12}^\nu - c_{23}^\nu c_{12}^\nu  \theta_{12}^e \mathrm{e}^{- \mathrm{i} \delta_{12}^e} \;,
\end{split}
\end{equation}
where we have already discarded all trivial phases and RG corrections which we will discuss later. For the total leptonic  mixing angles we obtain
\begin{equation}\label{eq:prediction}
\begin{split}
\theta_{12}^{\mathrm{MNS}}  & \approx 35.1^\circ \;, \\
\theta_{13}^{\mathrm{MNS}}  & \approx   3.3^\circ \;, \\
\theta_{23}^{\mathrm{MNS}}  & = 45.0^\circ\;.
\end{split}
\end{equation}
For the phases we have $\delta^{\mathrm{MNS}}_{13} = \pi - \delta_{12}^e \approx  94.6^\circ$, $\delta^{\mathrm{MNS}}_{12} = 4.6^\circ$ and $\delta^{\mathrm{MNS}}_{23} = 0^\circ$ from which the final MNS phases can be calculated according to \cite{Antusch:2008yc}
\begin{equation}\label{eq:prediction_phases}
\begin{split}
\delta_{\mathrm{MNS}} &= \delta_{13}^{\mathrm{MNS}} - \delta^{\mathrm{MNS}}_{12} \approx  90.0^\circ \;,\\
\alpha_1 &= 2 (\delta_{12}^{\mathrm{MNS}} + \delta_{23}^{\mathrm{MNS}}) = 2 \delta_{12}^{\mathrm{MNS}} \approx 9.3^\circ \;,\\
\alpha_2 &= 2 \delta_{23}^{\mathrm{MNS}} = 0^\circ \;,
\end{split}
\end{equation}
where $\alpha_1$ and $\alpha_2$ are the Majorana phases as in the PDG parameterization where they are contained in a diagonal matrix $\mathrm{diag}(\mathrm{e}^{\mathrm{i} \alpha_1/2}, \mathrm{e}^{\mathrm{i} \alpha_2/2}, 1)$.

We note that with the mixing pattern of our model, i.e.\ tri-bimaximal mixing produced in the neutrino sector, and charged lepton mixing corrections only from $\theta_{12}^e$, the leptonic mixing angles and the Dirac CP phase $\delta_{\mathrm{MNS}}$ satisfy the lepton mixing sum rule \cite{sumrule}
\begin{eqnarray}
\theta_{12}^{\mathrm{MNS}} - \theta_{13}^{\mathrm{MNS}} \cos (\delta_{\mathrm{MNS}}) \approx \arcsin (1/\sqrt{3}) \:.
\end{eqnarray}
The approximately maximal CP violation, i.e.\ $\delta_{\mathrm{MNS}} \approx 90^\circ$, affects that although the charged lepton corrections generate $\theta_{13}^{\mathrm{MNS}} \approx 3.3^\circ$, the solar mixing angle remains very close to its tri-bimaximal value of $\arcsin (1/\sqrt{3}) $. So far, we have discussed the neutrino mixing parameters at the GUT scale. To calculate the low energy values we have to take RG running of the parameters into account.

\subsection{Renormalisation Group Corrections}
For a quasi-degenerate neutrino mass spectrum, RG corrections to the neutrino parameters can in principle change the high scale values dramatically. However, as has been discussed for type II upgraded seesaw models in \cite{Antusch:2004xd} and more generally in \cite{Antusch:2003kp,Antusch:2005gp},
for small $\tan \beta$ and small neutrino Yukawa couplings (in our example model they are much smaller than $y_\tau$) the corrections to the mixing angles and CP phases are under control.
Setting the small Majorana phases to zero and with $\delta_{\mathrm{MNS}} \approx 90^\circ$, we can estimate in leading order \cite{Antusch:2003kp,Antusch:2005gp}
\begin{eqnarray}
\frac{\mathrm{d} \, \theta^{\mathrm{MNS}}_{12}}{\mathrm{d} \ln (\mu/\mu_0)}  &\approx&
- \frac{y_\tau^2}{32 \pi^2} \sin (2 \theta^{\mathrm{MNS}}_{12}) (s_{23}^{\mathrm{MNS}})^2  \frac{|m_1 + m_2|^2}{\Delta m^2_{\mathrm{sol}}}\:,\\
\frac{\mathrm{d} \, \theta^{\mathrm{MNS}}_{13}}{\mathrm{d} \ln (\mu/\mu_0)}  &\approx& 0 \:,\\
\frac{\mathrm{d} \, \theta^{\mathrm{MNS}}_{23}}{\mathrm{d} \ln (\mu/\mu_0)}  &\approx&
- \frac{y_\tau^2}{32 \pi^2} \sin (2 \theta^{\mathrm{MNS}}_{23})  \frac{(c_{12}^{\mathrm{MNS}})^2  |m_2+ m_3|^2 + (s_{12}^{\mathrm{MNS}})^2  |m_1+ m_3|^2 }{\Delta m^2_{\mathrm{atm}}} \:,
\end{eqnarray}
where $\mu$ is the renormalisation scale.
In the case of quasi-degenerate neutrino masses we can further use the approximation $m_3 \approx m_2 \approx m_1 = m_0 $. Integrating these equations approximately with the parameters on the right side taken constant and equal to their GUT scale values, and plugging in these numbers, we obtain the estimated low energy values of the mixing angles as
\begin{eqnarray}
\theta^{\mathrm{MNS}}_{12}|_{m_t(m_t)} &\approx& \theta^{\mathrm{MNS}}_{12}|_{M_{\mathrm{GUT}}} + 0.15^\circ \frac{m_0^2}{(0.1 \:\mbox{eV})^2} \:,\\
\theta^{\mathrm{MNS}}_{13}|_{m_t(m_t)} &\approx& \theta^{\mathrm{MNS}}_{13}|_{M_{\mathrm{GUT}}}  \:,\\
\theta^{\mathrm{MNS}}_{23}|_{m_t(m_t)} &\approx& \theta^{\mathrm{MNS}}_{23}|_{M_{\mathrm{GUT}}} \pm 0.01^\circ \frac{m_0^2}{(0.1  \:\mbox{eV})^2} \:.
\end{eqnarray}
In the last equation, the ``+'' applies for a normal neutrino mass ordering, whereas the ``-'' applies for an inverse mass odering, i.e. the case $\Delta m^2_{\mathrm{atm}} < 0$. It is important to note that both mass orderings can be realised in our model.
The strong suppression for the RG running of $\theta^{\mathrm{MNS}}_{13}$ is caused by the particular values of the CP violating phases in our model. For similar reasons the running of the CP phases themselves is also suppressed, as can be seen using the analytical results in \cite{Antusch:2003kp,Antusch:2005gp}. In summary, RG corrections are under control in our setup and only cause comparatively small corrections to the mixing parameters in the lepton sector.

In summary, the leptonic mixing parameters in our model are compatible with the experimental
1$\sigma$ ranges at low energy which are: $\theta_{12}^{\mathrm{MNS}} = (34.5\pm 1.0)^\circ$, $\theta_{13}^{\mathrm{MNS}} =  (5.7^{+3.0}_{-3.9})^\circ$ and $\theta_{23}^{\mathrm{MNS}} = (42.3^{+5.3}_{-2.8})^\circ$, taken from \cite{GonzalezGarcia:2010er}, as long as $m_0$ is smaller than the cosmological bounds suggest, $m_0 \lesssim 0.2$~eV \cite{Jarosik:2010iu}.

The values for the leptonic mixing angles and Dirac CP phase $\delta_{\mathrm{MNS}}$, resulting from our assumed vacuum alignment and stated in Eqs.~\eqref{eq:prediction} and \eqref{eq:prediction_phases} can be tested accurately by ongoing and future precision neutrino oscillation experiments \cite{Bandyopadhyay:2007kx}.

\subsection{Predictions for Beta Decay Experiments}

The effective mass relevant for neutrinoless double beta decay is
\begin{equation}\label{eq:mee_formula}
m_{ee}  = \vert m_1 c_{12}^2 c_{13}^2 e^{\mathrm{i} \alpha_1} + m_2 s_{12}^2 c_{13}^2 e^{\mathrm{i} \alpha_2} + m_3 s_{13}^2 e^{2 \mathrm{i} \delta_{\mathrm{MNS}}} \vert \;,
\end{equation}
while the kinematic mass accessible in the single beta decay end-point experiment KATRIN is
\begin{equation}
m^2_\beta \equiv m_1^2 c_{12}^2 c_{13}^2 + m_2^2 s_{12}^2 c_{13}^2  + m_3^2 s_{13}^2 \;.
\end{equation}
For quasi-degenerate neutrino mass spectrum ($m_0 = m_1 \simeq m_2 \simeq m_3$) we obtain that
\begin{equation}
m_\beta \approx m_0 \quad
\end{equation}
directly gives information about the absolute neutrino mass scale.

On the other hand, due to the phases appearing in Eq.~\eqref{eq:mee_formula} there is typically a sizable ambiguity in the relation between $m_{ee}$ and $m_0$, as long as the Majorana CP phases are not predicted.
Allowing, for instance, for arbitrary Majorana phases and considering a quasi-degenerate neutrino mass spectrum ($m_0 = m_1 \simeq m_2 \simeq m_3$) and with small $\theta^{\mathrm{MNS}}_{13}$, $m_{ee}$ can still be in the approximate interval $m_{ee} \in[m_{\mathrm{lightest}}/3,m_{\mathrm{lightest}}]$.

\begin{figure}
\centering
\includegraphics[scale=0.6]{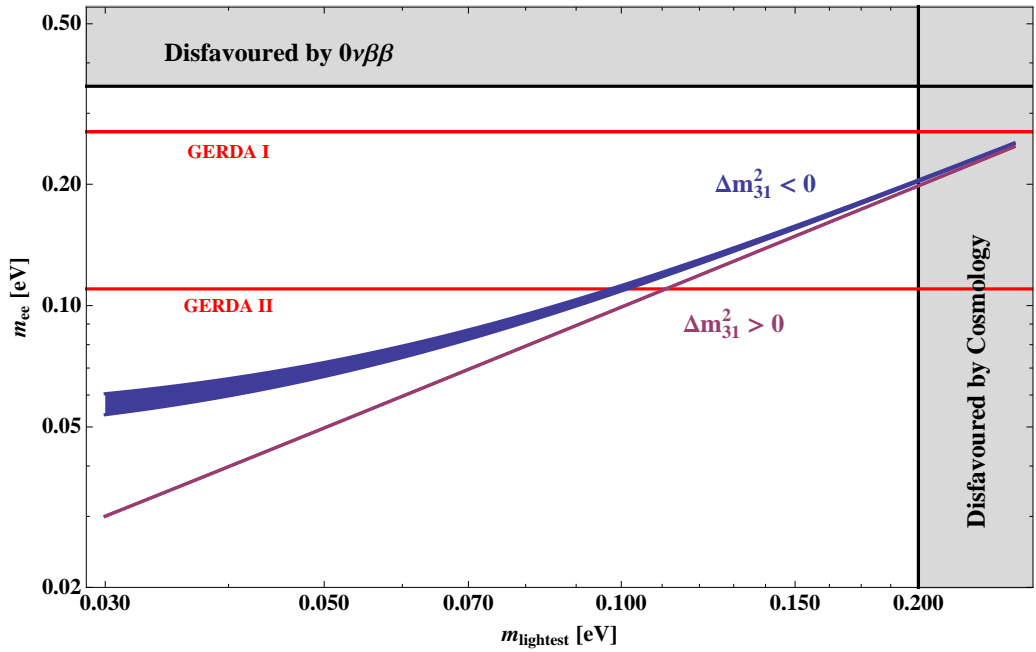}
\caption{The effective mass $m_{ee}$ in our setup relevant for neutrinoless double beta decay as a function of the mass $m_\mathrm{lightest}$ of the lightest neutrino, for an inverted neutrino mass ordering ($\Delta m_{31}^2 < 0$, upper line) and for a normal mass ordering ($\Delta m_{31}^2 > 0$, lower line). The bands represent the experimental uncertainties of the mass squared differences. The mass bounds from cosmology \protect\cite{Jarosik:2010iu} and from the Heidelberg-Moscow experiment \protect\cite{KlapdorKleingrothaus:2000sn} are displayed as grey shaded regions. The red lines show the expected sensitivities of the  GERDA experiment in phase I and II \protect\cite{Smolnikov:2008fu}.  \label{Fig:meePlot}}
\end{figure}

This ambiguity is resolved in our model, 
since the large contribution to the neutrino mass matrix proportional to the unit matrix\footnote{In our model this part of the neutrino mass matrix is induced by a standard type I (with right-handed neutrinos) and an additional type II seesaw contribution. Nevertheless, the conclusions remain the same as for the pure type II seesaw case since we assume that the type II contribution is dominant.} (with $\vert m_0 \vert \gg \vert m_3^I \vert$,~$\vert m_2^I \vert$) results in small Majorana CP phases and thus we predict:
\begin{equation}
m_{ee} \approx m_0 \;.
\end{equation}
The assumed dominance of $m_0$ in our model allows to realise a quasi-degenerate neutrino spectrum (with normal or inverse mass ordering) in a natural way, without any tuning of parameters.
The possible values for $m_{ee}$ as function of the lightest neutrino mass $m_{\mathrm{lightest}}$ is shown in Fig.~\ref{Fig:meePlot}.

We would like to remark that for smaller $m_0$ one can also naturally extend the model to hierarchical or inverted hierarchical neutrino masses without changing the leptonic mixing angles. With $\vert m_0 \vert \approx \vert m_2^I \vert$ or $\vert m_0 \vert \approx \vert m_3^I \vert$ or both we then also encounter cases where the Majorana phases are close to $\pi$. For a quasi-degenerate spectrum the model disfavours these unnatural cases since they would correspond to heavily fine-tuned parameters of the model. Similarly, an inverse strongly hierarchical spectrum would require unnatural tuning between $m_0$ and $m_3^I$ to make $m_3 = |m_0 - m_3^{I}|$ very small. By contrast, for a typical parameter choice of the model, a normally ordered hierarchical spectrum simply corresponds to $\vert m_0 \vert \ll \vert m_2^I \vert$,~$\vert m_3^I \vert$ and does not require any tuning at all.
It is also interesting to note that with the phases in our model there is {\em no} possibility to have cancellations in Eq.~\eqref{eq:mee_formula} that could make $m_{ee}$ vanish exactly.\footnote{In fact we find numerically that $m_{ee} \gtrsim 0.007$ eV.} Neutrinoless double beta decay is thus, also for smaller $m_0$, an unavoidable consequence in this class of models.

\section{Summary and Conclusions}

We have proposed a model of quasi-degenerate neutrinos
which predicts the neutrinoless double beta decay mass observable to be
approximately equal to the neutrino mass scale $m_{\mathrm{lightest}}$,
thereby allowing its determination approximately independently of unknown
Majorana phases.
In general such quasi-degenerate neutrino masses may
be naturally realised if there exists a non-Abelian family symmetry with real triplet
representations that enforces an additional contribution to the neutrino mass matrix proportional to the unit matrix,
hence determining the neutrino mass scale. In our model, the additional contribution was generated by a type II seesaw, or,  alternatively, by another type I seesaw contribution from an additional triplet representation of right-handed neutrinos (or neutrino messenger fields). 
In addition, the standard type I seesaw contribution determines the neutrino mixing angles.
Although such a mechanism (called a ``type II upgrade'' in \cite{Antusch:2004xd}) has been known for some time, the model in this paper is the first of its kind to combine this mechanism
with tri-bimaximal mixing arising from $A_4$ family symmetry together with
a $SU(5)$ SUSY GUT.

The SUSY $A_4\times SU(5)$ model considered here has several attractive features.
The full renormalisable superpotential for the coupling of matter to flavons, messenger 
fields and Higgs fields is specified, and only discrete auxiliary symmetries are
introduced, rather than the more common continuous Abelian symmetry that is typically invoked in such models.
In the considered model, the neutrino mass scale can either originate from the induced vacuum expectation value of a $SU(2)_\mathrm{L}$ triplet contained in a fifteen-dimensional representation of $SU(5)$, or from an additional contribution to the neutrino mass operator induced by $SU(5)$ singlet messenger fields in the triplet representation of $A_4$. Since the type I seesaw contribution involves very hierarchical additional neutrino mass contributions,
responsible for the tiny mass splittings between the quasi-degenerate neutrinos, we use the
constrained sequential dominance mechanism which is well suited for achieving such strongly
hierarchical contributions. In such a model the neutrino flavour symmetry associated with
tri-bimaximal mixing is achieved indirectly from the family symmetry, given the assumed
vacuum alignment.
We have included renormalisation group corrections to the mixing angles and shown that they are
under control for small values of $\tan \beta$ in our framework and do not significantly modify the leptonic mixing angles.
We have therefore considered small values of $\tan \beta$ throughout our study.

The model has several interesting phenomenological features which emerge from our numerical fit to the quark and lepton masses and quark mixing. In addition to $b$-$\tau$ unification that may be viable for small $\tan \beta$, the model realises the GUT scale relation $y_\mu/y_s \approx 9/2$ proposed in \cite{Antusch:2009gu}, which is favourable for small values of $\tan \beta$. In the quark sector, we observe that the correct $\delta_{\mathrm{CKM}}$ (corresponding to the right unitarity triangle with $\alpha \approx 90^\circ$) can be realised for the simple ansatz in Eq.~\eqref{ansatz}. We remark that this simple ansatz leads to an interesting alternative texture (different from \cite{Antusch:2009hq}) that gives rise to 
a  right-angled unitarity triangle with $\alpha\approx 90^\circ$. For the leptonic mixing angles and CP phases we find $\theta_{13}^{\mathrm{MNS}} \approx 3.3^\circ$, $\theta_{12}^{\mathrm{MNS}} \approx 35.1^\circ$, $\theta_{23}^{\mathrm{MNS}} \approx 45.0^\circ$ and $\delta_{\mathrm{MNS}} \approx 90^\circ$. The leptonic mixing angles satisfy the ``lepton mixing sum rule'' proposed in \cite{sumrule} with only small theoretical errors since the 1-3 mixing in the charged lepton mass matrix is very small and the 2-3 mixing vanishes. The Majorana CP phases are small for quasi-degenerate neutrino masses via a large additional contribution proportional to the unit matrix, as expected, and the model thus predicts the neutrinoless double beta decay mass observable to be approximately equal to the neutrino mass scale, or lightest neutrino mass, i.e.\  $m_{ee} \approx m_{\mathrm{lightest}}$.

In conclusion, if neutrinoless double beta decay were observed in the near future then this would
herald another neutrino revolution in which neutrino masses would be quasi-degenerate.
Amongst the many possible models of quasi-degenerate neutrinos, the model with an additional contribution to neutrino mass matrix proportional to the unit matrix, as considered here,
are distinguished by their prediction of that the neutrinoless double beta decay mass observable is approximately equal to the neutrino mass scale $m_{\mathrm{lightest}}$.
We have proposed the first realistic model of this kind involving $A_4$ and SUSY $SU(5)$ GUTs.
The $A_4$ family symmetry has the dual effect of enforcing on the one hand
quasi-degeneracy via the additional unit matrix contribution to the neutrino mass matrix,  
and on the other hand tri-bimaximal mixing via the type I seesaw mechanism and constrained sequential dominance.
A numerical fit to quark masses and mixing angles reveals that
a simple ansatz describing quark CP violation with
$\alpha \approx 90^\circ$ also leads to
the leptonic phase  $\delta_{\mathrm{MNS}} \approx 90^\circ$.
In such models the absolute neutrino masses could be directly measurable by experiment quite soon.

\section*{Acknowledgments}
We would like to thank Michal Malinsk\'{y} for valuable discussions, especially
during the initial stages of the project, and Christoph Luhn for valuable discussions on possibilities for the vacuum alignment. 
We also thank Vinzenz Maurer for very valuable comments on the manuscript.
We are indebted to NORDITA for the hospitality and support during the programme
``Astroparticle Physics -- A Pathfinder to New Physics'' held in
Stockholm in March 30 - April 30, 2009 during which part of this
study was performed.
S.~A.\ and M.~S.\ acknowledge partial support by the DFG cluster of excellence
``Origin and Structure of the Universe.''
S.~F.~K.\ acknowledges partial support from the STFC
Rolling Grant No. ST/G000557/1 and a Royal Society Leverhulme Trust Senior Research Fellowship.

\appendix
\section{The Renormalisable Superpotential and Effective Couplings}

With the field content and symmetries specified in Tab.~\ref{Tab:Symmetries} the superpotential
contains the following renormalisable terms:
\begin{align}
W_{H} &= \mu_{5} H_5 \bar{H}_5 + \mu_{15} H_{15} \bar{H}_{15} + \mu_{45} H_{45} \bar{H}_{45}+ \bar{\lambda}_{15} \bar{H}_{15}  H_5 H_5 + \lambda_{15} H_{15} \bar{H}_5 \bar{H}_5 \;, \label{Eq:mu} \\
W_A &= M_{A_{10}} A_{10} \bar{A}_{10} + M_{A_5} A_5 \bar{A}_5 + M_{A_1} A^2_1 + M_B B \bar{B}  + M_{C_1} \bar{C}_1 C_1 + M_{C_2} \bar{C}_2 C_2 \label{Eq:Xmass} \\
W_{\mathrm{int}} &=  \kappa_{F} F \bar{H}_5 A_{10} + \tilde{\kappa}_{F} F \tilde{\phi}_{23} A_5 + \kappa_{Ti} T_i \phi_i \bar{A}_{10} + \tilde{\kappa}_{T2} T_2 \bar{H}_{45} \bar{A}_5 + y_\Delta H_{15} F F \nonumber\\
& + \lambda H_5 A_{10} \bar{A}_{10} + \tilde{\lambda} H_5 \bar{C}_2 \bar{B} + \kappa'_T B T_2^2 + \kappa'_{\phi} C_2 \phi_{123}^2 + \tilde{\kappa}'_{\phi} C_2 \tilde{\phi}_{23}^2 + a_t H_5 T_3^2 \nonumber\\
& + \bar{\kappa}_{F} F \bar{H}_5 A_1 + \kappa_{N i} N_i \phi_i A_1 + \xi_{\phi i} C_i \phi_i^2 + \xi_{N i} \bar{C}_i N_i^2 \label{Eq:Wint} \;.
\end{align}

As discussed in the main text,
after GUT symmetry breaking the $SU(2)_L$ doublet components from $H_5$ and $H_{45}$ respectively $\bar{H}_5$ and $\bar{H}_{45}$  mix and only the light states acquire the $SU(2)_L$ breaking vevs which give the fermion masses.
We parameterise the Higgs mixing with the mixing angles $\gamma$ respectively $\bar{\gamma}$
\begin{equation}
\begin{split}
\begin{pmatrix} H_5 \\ H_{45} \end{pmatrix} &= \begin{pmatrix} c_\gamma & - s_\gamma \\  s_\gamma & c_\gamma \end{pmatrix} \begin{pmatrix} H_l \\ H_{h} \end{pmatrix} \;, \\
\begin{pmatrix} \bar{H}_{5} \\ \bar{H}_{45} \end{pmatrix} &= \begin{pmatrix} c_{\bar{\gamma}} & - s_{\bar{\gamma}} \\  s_{\bar{\gamma}} & c_{\bar{\gamma}} \end{pmatrix} \begin{pmatrix} \bar{H}_l \\ \bar{H}_{h} \end{pmatrix}  \;,
\end{split}
\end{equation}
where we have used the common abbreviation $c_\gamma \equiv \cos \gamma$ and similar for the sine and the other angle $\bar{\gamma}$. The light Higgs doublets are denoted with an index $l$ while the heavy Higgs doublets are denoted with an index $h$.

The effective couplings $a$ appearing in the effective superpotential in the main text
can be expressed in term of the fundamental couplings from Eq.~\eqref{Eq:Wint} and the messenger masses from \eqref{Eq:Xmass}
\begin{equation}
\label{Eq:ais}
a_i =  \kappa_{F} \kappa_{Ti}  \;, \quad \tilde{a}_2 = \tilde{\kappa}_{F} \tilde{\kappa}_{Ti} \;,
\end{equation}
\begin{equation}
\begin{split}
a_{11} =  \lambda \kappa_{T 1}^2 \;, \quad & a_{22} = \lambda \kappa_{T 2}^2 + \tilde{\lambda} \kappa'_T \kappa'_\phi \frac{M_{A_{10}}^2 }{M_{B} M_{C_2}} \;, \quad a_{33} =  a_t +  \frac{\lambda \kappa_{T 3}^2}{M_{A_{10}}^2 } \;,\\
& \tilde{a}_{22} = \tilde{\lambda} \kappa'_T  \tilde{\kappa}'_{\phi} \frac{M_{A_5}^2}{M_{B} M_{C_2} } \;, \quad  \quad a_{ij} = \lambda \kappa_{T i} \kappa_{T j} \;, \text{for $i \neq j$} \;,
\end{split}
\end{equation}
\begin{equation}
a_{\nu_i} = \kappa_{Ni} \bar{\kappa}_F \frac{M_{A_{10}}}{M_{A_1}} \;,
\end{equation}
\begin{equation}
\label{Eq:aRs}
a_{R_i}= \kappa_{N i}^2 \frac{M_{A_{10}}^2}{M_{A_1}} + \xi_{\phi_i} \xi_{N i} \frac{M_{A_{10}}^2}{M_{C_i}}  \;,\quad a_{R_{12}} = \kappa_{N i} \kappa_{N j} \frac{M_{A_{10}}^2}{M_{A_1}} \;.
\end{equation}

The effective couplings $b$ appearing in the Yukawa couplings can be expressed in terms of the couplings $a$ and the Higgs mixing angles as
\begin{equation}
\label{Eq:bis}
b_i =  c_{\bar{\gamma}} a_i   \;, \quad \tilde{b}_2 = s_{\bar{\gamma}} \tilde{a}_{2}  \;,
\end{equation}
\begin{equation}
b_{ij} =  \frac{ c_{\gamma} }{c^2_{\bar{\gamma}}} \frac{ a_{ij} }{ a_i a_j }   \;, \quad \tilde{b}_{22} = \frac{ c_{\gamma} }{s^2_{\bar{\gamma}} }  \frac{ \tilde{a}_{22} }{\tilde{a}_2^2}  \;, \quad b_{33} = c_{\gamma} a_t + \frac{ c_{\gamma} }{c^2_{\bar{\gamma}}}  \frac{ \epsilon_3^2 }{a_3^2} \;,
\end{equation}
\begin{equation}
b_{\nu_i} = \frac{c_\gamma}{ c_{\bar{\gamma}}} \frac{ a_{\nu_i} }{a_i} \;,
\end{equation}
\begin{equation}
\label{Eq:bRs}
b_{R_i}= \frac{1}{c_{\bar{\gamma}}^2} \frac{ a_{R_i} }{a_i^2 } \;,\quad b_{R_{12}} = \frac{1}{c_{\bar{\gamma}}^2} \frac{ a_{R_{12}} }{a_1 a_2}\;.
\end{equation}

\section{A Possible Vacuum Alignment}

\begin{table}
\centering
\begin{tabular}{ccccccccc}
\toprule
& $SU(5)$ & $A_4$ & $\mathbb{Z}_2$ & $\mathbb{Z}^{(1)}_4$ & $\mathbb{Z}^{(2)}_4$ & $\mathbb{Z}^{(3)}_4$ & $\mathbb{Z}^{(4)}_4$ & $U(1)_R$\\
\midrule
\multicolumn{8}{l}{Chiral Matter}  \\
\midrule
$F$ &  $\mathbf{\overline{5}}$ & $\mathbf{3}$ & + & 0 & 0 & 0 & 0 & 1 \\
$T_1$, $T_2$, $T_3$  &  $\mathbf{10}$, $\mathbf{10}$, $\mathbf{10}$ &  $\mathbf{1}$, $\mathbf{1}$, $\mathbf{1}$ & +, +, - &  0, 1, 0 & 0, 0, 0 &  0, 0, 0 &  1, 0, 0 & 1, 1, 1  \\
$N_1$, $N_2$ &  $\mathbf{1}$, $\mathbf{1}$ &  $\mathbf{1}$, $\mathbf{1}$ & +, + &  0, 1 & 0, 0 &  0, 0 &  1, 0  & 1, 1  \\
\midrule
\multicolumn{8}{l}{Flavons}  \\
\midrule
$\phi_{23}$, $\phi_{123}$, $\phi_{3}$ & $\mathbf{1}$, $\mathbf{1}$, $\mathbf{1}$ & $\mathbf{3}$, $\mathbf{3}$, $\mathbf{3}$  & +, +, - & 0, 3, 0 & 0, 0, 0 & 0, 0, 0 & 3, 0, 0 & 0, 0, 0 \\
$\tilde{\phi}_{1}$, $\tilde{\phi}_{2}$, $\tilde{\phi}_{3}$ & $\mathbf{1}$, $\mathbf{24}$, $\mathbf{1}$ & $\mathbf{3}$, $\mathbf{3}$, $\mathbf{3}$   & -, +, + & 3, 3, 3 & 3, 3, 0  & 3, 0, 3 & 0, 0, 0   & 0, 0, 0\\
$\tilde{\xi}$ &  $\mathbf{1}$, $\mathbf{1}$ & $\mathbf{1}$, $\mathbf{1}$ & + & 1, 3 & 0, 0 & 0, 0 & 0, 0 & 0, 0\\
\midrule
\multicolumn{8}{l}{Higgs Multiplets}  \\
\midrule
$H_5$, $\bar{H}_5$ &  $\mathbf{5}$, $\mathbf{\overline{5}}$ & $\mathbf{1}$, $\mathbf{1}$ & +, + & 0, 0 & 0, 0 & 0, 0 & 0, 0 & 0, 0\\
$H'_5$, $\bar{H}'_5$ &  $\mathbf{5}$, $\mathbf{\overline{5}}$ & $\mathbf{1}$, $\mathbf{1}$ & +, + & 0, 0 & 0, 0  & 3, 1 & 0, 0 & 0, 0\\
$H_{15}$, $\bar{H}_{15}$ &  $\mathbf{15}$, $\mathbf{\overline{15}}$ &  $\mathbf{1}$, $\mathbf{1}$  & +, + &  0, 0 & 0, 0 &  0, 0 & 0, 0 & 0, 0\\
$H_{45}$, $\bar{H}_{45}$ &  $\mathbf{45}$, $\mathbf{\overline{45}}$ &  $\mathbf{1}$, $\mathbf{1}$  & +, + &  0, 0 & 3, 1 &  0, 0 & 0, 0 & 0, 0\\
\midrule
\multicolumn{8}{l}{Matter-like Messengers } \\
\midrule
$A_5$, $\bar{A}_5$ &  $\mathbf{5}$, $\mathbf{\overline{5}}$ & $\mathbf{1}$, $\mathbf{1}$ & +, + & 1, 3 & 1, 3 & 0, 0 &  0, 0 & 1, 1 \\
$A'_5$, $\bar{A}'_5$ &  $\mathbf{5}$, $\mathbf{\overline{5}}$ & $\mathbf{1}$, $\mathbf{1}$ & +, + &   1, 3 & 0, 0 & 1, 3 &  0, 0 & 1, 1 \\
$A_{10}$, $\bar{A}_{10}$ &  $\mathbf{10}$, $\mathbf{\overline{10}}$ & $\mathbf{3}$, $\mathbf{3}$ & +, + & 0, 0 & 0, 0 &  0, 0 &  0, 0   & 1, 1 \\
$A_{1}$ &  $\mathbf{1}$ & $\mathbf{3}$ & + & 0 & 0 & 0 & 0 & 1\\
\midrule
\multicolumn{8}{l}{Higgs-like Messengers }  \\
\midrule
$B$, $\bar{B}$ & $\mathbf{5}$, $\mathbf{\overline{5}}$  & $\mathbf{1}$, $\mathbf{1}$ & +, + & 2, 2 & 0, 0 & 0, 0 &  0, 0  & 0, 2  \\
$C_{23}$, $\bar{C}_{23}$ & $\mathbf{1}$, $\mathbf{1}$  & $\mathbf{1}$, $\mathbf{1}$ & +, + &  0, 0 & 0, 0  & 0, 0 &  2, 2 & 2, 0  \\
$C_{123}$, $\bar{C}_{123}$ & $\mathbf{1}$, $\mathbf{1}$  & $\mathbf{1}$, $\mathbf{1}$ & +, + &  2, 2 & 0, 0 & 0, 0 &  0, 0 & 2, 0  \\
$C_1$, $\bar{C}_1$ & $\mathbf{1}$, $\mathbf{1}$  & $\mathbf{1}$, $\mathbf{1}$ & +, + &  2, 2 & 2, 2 & 2, 2 &  0, 0 & 2, 0  \\
$C_2$, $\bar{C}_2$ & $\mathbf{1}$, $\mathbf{1}$  & $\mathbf{1}$, $\mathbf{1}$ & +, + &  2, 2 & 2, 2 & 0, 0 &  0, 0 & 2, 0  \\
$C_3$, $\bar{C}_3$ & $\mathbf{1}$, $\mathbf{1}$  & $\mathbf{1}$, $\mathbf{1}$ & +, + &  2, 2 & 0, 0 & 2, 2 &  0, 0 & 2, 0  \\
\midrule
\multicolumn{8}{l}{Driving Fields}  \\
\midrule
$\tilde{P}_i$, $P_i$ & $\mathbf{1}$, $\mathbf{1}$ & $\mathbf{1}$, $\mathbf{1}$ & +, + & 0, 0 & 0, 0 & 0, 0 & 0, 0 & 2, 2 \\
$D_{123}$, $D_3$ & $\mathbf{1}$, $\mathbf{1}$ & $\mathbf{3}$, $\mathbf{3}$ & +, + & 2, 0 & 0, 0 & 0, 0 & 0, 0 & 2, 2 \\
$\tilde{D}_1$, $\tilde{D}_2$, $\tilde{D}_3$  & $\mathbf{1}$, $\mathbf{1}$, $\mathbf{1}$ & $\mathbf{3}$, $\mathbf{3}$, $\mathbf{3}$ & +, +, + & 2, 2, 2 & 2, 2, 0 &  2, 0, 2 & 0, 0, 0 & 2, 2, 2 \\
$\tilde{O}_{12}$, $\tilde{O}_{13}$, $\tilde{O}_{23}$ & $\mathbf{24}$, $\mathbf{1}$, $\mathbf{24}$ & $\mathbf{1}$, $\mathbf{1}$, $\mathbf{1}$ & -, -, + & 2, 2, 2 & 2, 1, 1 &  1, 2, 1 & 0, 0, 0 & 2, 2, 2 \\
$O_{13}$, $O_{23}$ & $\mathbf{1}$, $\mathbf{24}$ & $\mathbf{1}$, $\mathbf{1}$ & -, - & 1, 1 & 1, 1 &  1, 0 & 0, 0 & 2, 2 \\
$O_{1;23}$, $O_{123;23}$ & $\mathbf{1}$, $\mathbf{1}$ & $\mathbf{1}$, $\mathbf{1}$ & -, + & 1, 1 & 1, 0 &  1, 0 & 0, 1 & 2, 2 \\
\bottomrule
\end{tabular}
\caption{Representations and charges of the superfields. The subscript $i$ on the fields $T_i$ and $N_i$ is a family index while for $\tilde{P}_i$ and $P_i$ the $i$ matches the corresponding flavon field.  The flavon fields $\phi_i$, $\tilde{\phi}_{23}$ can be associated to a family via their charges under $\mathbb{Z}_2 \times \mathbb{Z}_4^4$.  The subscripts on the Higgs fields $H$, $\bar{H}$ and extra vector-like matter fields $A$, $\bar{A}$ denote the transformation properties under $SU(5)$. \label{Tab:Symmetries2}}
\end{table}

In this section of the appendix, we will discuss a possibility to extend the model in the main part by a viable vacuum alignment. This possibility is based on realising the flavon  $\tilde{\phi}_{23}$ effectively by splitting it up into two flavons $\tilde{\phi}_{2}$ and $\tilde{\phi}_{3}$, which get purely real or purely imaginary vevs by a vacuum alignment as described below, i.e.:
 \begin{equation}
\langle \tilde{\phi}_{23} \rangle =
\langle \tilde{\phi}_2 \rangle + \langle \tilde{\phi}_3 \rangle \quad\text{where}\quad 
\langle \tilde{\phi}_2 \rangle  = \begin{pmatrix} 0 \\ -\text{i}  \\ 0  \end{pmatrix} \tilde{\epsilon}_{23} \quad\text{and}\quad \langle \tilde{\phi}_3 \rangle  = \begin{pmatrix} 0 \\ 0 \\ w \end{pmatrix} \tilde{\epsilon}_{23} \; .
\end{equation}
For the effective superpotential in Eqs.~(\ref{Eq:Yl}) - (\ref{Eq:MR}) this would mean to simply replace $\tilde{\phi}_{23}^2 \to \tilde{\phi}_{2}^2 + \tilde{\phi}_{3}^2$ and $\tilde{\phi}_{23} \bar H_{45} \to \tilde{\phi}_{2} \bar H_{45}  + \tilde{\phi}_{3} \bar H'_{5}$, with an additional Higgs field $H'_{5}$. The field content and the symmetries of this extended version of the model is given in Tab.\ \ref{Tab:Symmetries2}. We note that, at the level of precision discussed here, the model predictions are the same as in the main part of the paper. 

We now turn to the discussion of the vacuum alignment sector with the desired purely real or imaginary alignment (see also \cite{imagalign}). At the effective theory level the new operators involving, the ``driving fields'' $P_i$, $D_i$ and $O_i$, that are generated by the fields and symmetries specified in Tab.\ \ref{Tab:Symmetries2} are given by (dropping ${\cal O}(1)$ coupling constants):
\begin{align}
W_{\tilde{\phi}} &=\tilde{P}_i \left( \frac{\tilde{\phi}_i^4}{M^2_{C_i}} - \tilde{M}_i^4 \right)   + \tilde{D}_i (\tilde{\phi}_i \star \tilde{\phi}_i) + \tilde{O}_{ij} \tilde{\phi}_i \tilde{\phi}_j \;,\\
W_{\phi_{123}} &=  D_{123} (\tilde{\xi} \phi_{123} + \phi_{123} \star \phi_{123})  + P_{123} \left( \frac{\phi_{123}^4}{M^2_{C_{123}}}  + \frac{\tilde{\xi}^4}{M^2_{C_{123}}}  + \frac{\phi_{123}^2 \tilde{\xi}^2}{M^2_{C_{123}}}  - M_{123}^4  \right)   \;,\\
 W_{\phi_3} &= P_3 (\phi_3^2 - M_3^2 ) + D_3 (\phi_3 \star \phi_3) + O_{13} \tilde{\phi}_1 \phi_3 + O_{23} \tilde{\phi}_2 \phi_3 \;, \\
 W_{\phi_{23}} &= P_{23} \left( \frac{\phi_{23}^4}{M^2_{C_{23}}}  - M_{23}^4 \right)  + O_{1;23} \tilde{\phi}_1 \phi_{23} + O_{123;23} \phi_{123} \phi_{23} \;.
\end{align}

The minimisation of the resulting F-term potential leads to the desired vacuum alignment for all flavons. We note that the new flavons $\tilde\xi$ and $\tilde{\phi}_1$ are only ``auxiliary'' in the sense that they do not couple to the matter sector but are only relevant for the flavon alignment. The flavon $\tilde{\xi}$ leads to an additional term contributing to the charm mass, however its effect can be absorbed in the messenger masses such that the predictions of the model are unchanged. The four parts $W_{\tilde{\phi}}$, $W_{\phi_{123}}$, $W_{\phi_3}$ and $W_{\phi_{23}}$ have the following meaning: 
 
\begin{itemize}
\item The terms in $W_{\tilde{\phi}}$ enforce non-vanishing vevs of the three flavons $\tilde{\phi}_{1}$, $\tilde{\phi}_{2}$ and $\tilde{\phi}_{3}$ that are orthogonal to each other and have only one nonzero element, due to the effects of the terms with the driving fields $\tilde{D}_i$. Here the $\star$ denotes the symmetric triplet combination of $A_4$ (see e.g.~\cite{King:2006np}). With real $M_i$ due to the assumed spontaneous CP violation $\tilde{\phi}_{i}^4$ are forced to be real which means that the vevs $\langle\tilde{\phi}_{i}\rangle$ are either purely real or imaginary. We will choose a vacuum where $\tilde{\phi}_{2}$ is imaginary and $\tilde{\phi}_{3}$ is real. The vev of $\tilde{\phi}_1$ can be either real or imaginary and since its phase does not affect the results.
\item The part $W_{\phi_{123}}$ leads to the desired alignment for the flavon vev $\langle{\phi}_{123}\rangle$ along the $(\pm 1,\pm 1,\pm 1)$ direction (see also \cite{Altarelli:2005yx}) in the minima of the potential with $\langle{\phi}_{123}\rangle\not=0$ and $\langle \tilde\xi \rangle\not=0$. Again, due to the ${\phi}_{123}^4$ in the term with $P_{123}$ the vev can be either real or imaginary, and we choose the real vacuum in the $(1,1,1)$ direction. 
\item The terms in $W_{\phi_3}$ and $W_{\phi_{23}}$ finally provide the alignment for the flavons $\phi_3$ and $\phi_{23}$. 
The terms with the driving fields $O_i$ force the vev of $\phi_3$ to be orthogonal to $\langle\tilde{\phi}_{1}\rangle$ and $\langle\tilde{\phi}_{2}\rangle$, and the vev of  $\phi_{23}$ to be orthogonal to $\langle\tilde{\phi}_{1}\rangle$ and $\langle{\phi}_{123}\rangle$, leading to the alignments in the desired direction in flavour space. We can again choose the purely real vacuum as explained above. 
\end{itemize}

We note that the driving fields $P_i$ and $\tilde{P}_i$ have the same quantum numbers and therefore, a priori, a linear combination of such fields couples to each of the above terms ${\tilde{\phi}_i^4}/{M^2_{C_i}} - \tilde{M}_i^4$ and ${\phi_{i}^4}/{M^2_{C_{i}}}  - M_{23}^4$. Without loss of generality we have written the driving fields here already in the field basis where only one driving field couples to one of these terms. This can always be achieved by a proper field redefinition.

\end{document}